\documentclass[preprint2,twoside]{hwo}

\usepackage{lipsum}

\input{hwo.h}

\begin{document}

\title{\textbf{\LARGE Combining a Diffraction-Limited Coronagraph with Fiber Nulling: A Demonstration of Serially Coupling Different Nullers
}}

\author {\textbf{\large Satoshi Itoh,$^{1}$ Taro Matsuo,$^{1,2}$ Reiki Kojima,$^1$ Motohide Tamura,$^{^{3,4,5}}$ Takahiro Sumi,$^2$ and Oliver Guyon,$^{4,5,6}$}}
\affil{$^1$\small\it Nagoya University, Nagoya, Aichi, Japan}
\affil{$^2$\small\it Osaka University, Toyonaka, Osaka, Japan}
\affil{$^3$\small\it University of Tokyo, Bunkyo, Tokyo, Japan}
\affil{$^4$\small\it Astrobiology Center, Mitaka, Tokyo, Japan}
\affil{$^5$\small\it National Astronomical Observatory of Japan, Mitaka, Tokyo, Japan}
\affil{$^6$\small\it University of Arizona, Tucson, Arizona, USA}

\begin{abstract}
We present experimental results of an efficient small-IWA ($\sim$1 $\lambda/D$) high contrast imaging approach realized by co-optimizing a coronagraph front-end with a fiber nulling 2nd stage. The setup includes the one-dimensional diffraction-limited coronagraph (1DDLC) and Parity Fiber Nuller (PFN). The 1DDLC has promising features (binary nuller, small inner working angles (IWAs)). Although the 1DDLC has the 2nd /4th–order sensitivity to spectral bandwidth and tilt aberrations,  it outputs stellar leak due to wavelengths other than the design wavelength only as a flat wavefront on the Lyot-stop plane, preserving the same complex amplitude profile as an on-axis point source. The PFN after the 1DDLC erases the leak from the 1DDLC. For the wavelength 6-\% less than the coronagraph's design-center wavelength, we confirmed the contrast mitigation ability of $3.5\times10^{-5}$, which is about 1/20 times the value of the case with only 1DDLC, suggesting that the combined system works robustly against the broad spectral bandwidth. Future work needs to address the demonstration of the anticipated broadband robustness for the contrast level lower than about $10^{-5}$.
  \\
  \\
\end{abstract}

\vspace{2cm}

\section{Introduction}
The Habitable Worlds Observatory (HWO) will enable spectral and temporal characterization of reflected light in the visible and near-infrared wavelengths from nearby potentially habitable planets \citep{2018AsBio..18..663S}.
In the HWO mission, a large number of sample planets means a precise statistical assessment, including evaluation of the life-emergence rate at potentially habitable planets \citep{2012PNAS..109..395S}.
In this context (i.e., increasing the number of observed planets), coronagraphic research may contribute to the enhancement of the HWO mission yields.

In this conference paper, we briefly report an experimental verification of a serially combined nuller \citep{2025PASP..137h4401I}.
The nuller theoretically has the ability to null the double stars simultaneously and diffraction-limited ($\sim$1 $\lambda/D$) inner working angles.
In Section 2, we briefly review the theory for components of the combined nuller and explain the design optimization \citep{2024AJ....167..235I} of the conjoined system.
Sections 3 describe the experimental result. 
Section 5 summarizes the present research.
\section{Theory}
\subsection{First-Stage Nuller}
\begin{figure}[htb]
    \centering
    \includegraphics[width=0.7\linewidth]{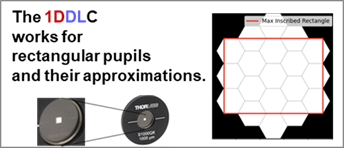}
    \caption{Pupil geometry suitable for the 1DDLC. The right panel shows an example of the inscribed rectangle for a possible HWO primary mirror geometry. The left picture shows the actual rectangle aperture used in the experiment. }
    \label{fig:1}
\end{figure}

In the present experiment, we use a type of Lyot coronagraph referred to as the one-dimensional diffraction-limited coronagraph (1DDLC).
The 1DDLC works well for the rectangular pupils and their approximation (Figure \ref{fig:1}), while most of the proposed Lyot coronagraphs work well for circular pupils.   
The derivation of the 1DDLC \citep{Itoh+2020} uses a simple one-dimensional analytic calculation similar to but different from the band-limited mask coronagraph \citep{2002ApJ...570..900K}.
\begin{figure*}[htb]
    \centering
    \includegraphics[width=0.7\linewidth]{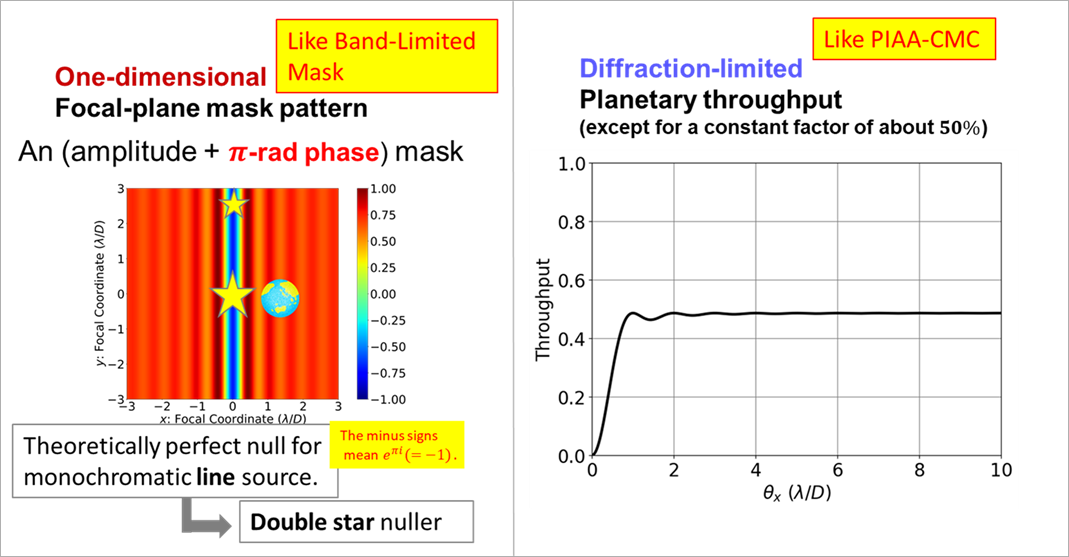}
    \caption{Two main characteristics of the 1DDLC.}
    \label{fig:2}
\end{figure*}

The 1DDLC has the following two promising natures (Figure \ref{fig:2}). (i) The focal-plane mask (FPM) takes constants along one direction. In other terms, the FPM function includes only one variable (a one-dimensional function). This nature enables to null the monochromatic line sources on the sky. Hence, the 1DDLC can erase the double stars simultaneously. (ii) The 1DDLC can achieve the diffraction-limited planetary throughput (i.e., $\sim$ 1 $\lambda/D$ inner working angles). 
For example, the band-limited mask coronagraph \citep{2002ApJ...570..900K} has a one-dimensional nature but non-diffraction-limited inner working angles.
The phase-induced amplitude apodization complex mask coronagraph \citep[PIAA-CMC,][]{Guyon+2010} has a diffraction-limited inner working angle but non-one-dimensional FPM.

We can write the mathematical expression for the FPM function $M(x)$ of the 1DDLC as follows:
\begin{equation}
    M(x)=s\left(1-2\mathrm{sinc(2x)}\right),
\end{equation}
where $x$ is the $x$-component of the focal cartesian coordinates $(x,y)$ normalized by $\lambda_c/D$ ($\lambda_c$: design center wavelength), 
\begin{equation}
    s=0.697...
\end{equation}
and
\begin{equation}
    \mathrm{sinc}(z)=\frac{\sin\left(\pi z\right)}{\pi z}.
\end{equation}
Here, we require the factor $s$ only to normalize the mask function $M(x)$ so that $\left|M(x)\right|\leq 1$, assuming no possibilities for focal-plane amplification of the light amplitude.
In simulation, we can utilize the built-in functions of programming tools, such as the NumPy module in Python and the MATLAB language, for the above sinc function $\mathrm{sinc}(z)$.  
The left panel of Figure \ref{fig:2} shows the two-dimensional colormap of the FPM function $M(x)$.

\begin{figure}[htb]
    \centering
    \includegraphics[width=0.6\linewidth]{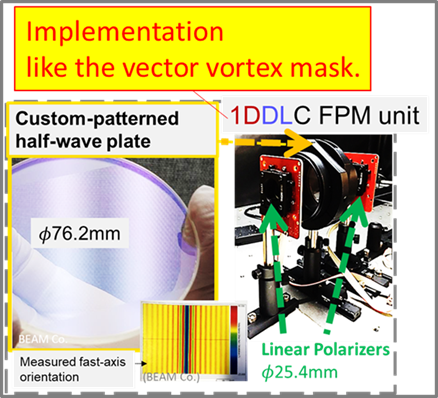}
    \caption{The implementation of the 1DDLC mask used in the experiment.}
    \label{fig:3}
\end{figure}

We can simply implement the FPM of the 1DDLC with a method (Figure \ref{fig:3}) similar to the vector vortex mask coronagraph \citep{2010ApJ...709...53M}.
In the case of the 1DDLC, we locate a custom-patterned half-wave plate between two linear polarizers.
This configuration can implement FPMs that take the values in the interval $[-1,1]$ on the real axis in the complex number plane.
Note that the vector vortex mask coronagraphs have a custom-patterned half-wave plate between two `circular' polarizers to implement the mask values on the unit circle in the complex number plane.

\begin{figure}[htb]
    \centering
    \includegraphics[width=1.0\linewidth]{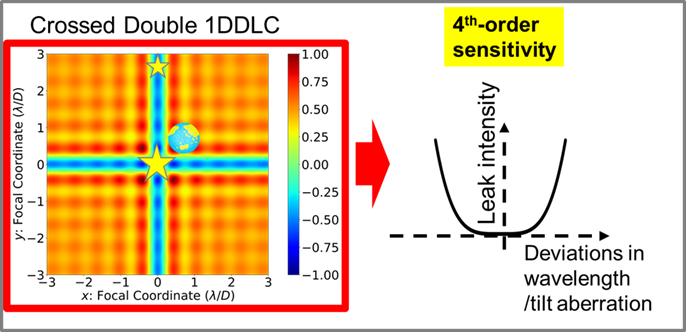}
    \caption{The crossed double 1DDLC. The left panel shows the FPM function of the crossed double 1DDLC.}
    \label{fig:4}
\end{figure}

The 1DDLC exhibits second-order sensitivities to wavelength deviations from the design-center wavelength and tilt aberrations.
In other words, regarding the deviation of wavelength from the design-center wavelength, the leakage in the 1DDLC is proportional to the value $\left(\frac{\lambda-\lambda_c}{\lambda_c}\right)^2$.
In a similar manner, when considering the tilt aberration from the stellar non-zero angular diameter, we observe the amount of the 1DDLC's leak proportional to the value $\left(\Delta \theta\right)^2$, where $\Delta \theta$ means the angular deviation from the optical axis normalized by $\lambda_c/D$.
The crossed double 1DDLC (Figure \ref{fig:4}) exhibits fourth-order sensitivities to wavelength deviations from the design-center wavelength and tilt aberrations \citep{Itoh+2020}. 
We can write the FPM function of the crossed double 1DDLC as follows:
\begin{equation}
    M_{\mathrm{cross}}(x,y)=M(x)M(y).
\end{equation}

\begin{figure}[htb]
    \centering
    \includegraphics[width=0.7\linewidth]{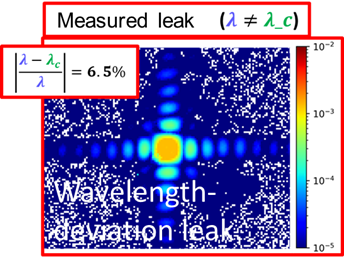}
    \caption{Measured focal intensity map of the 6.5-\% wavelength deviation leak. }
    \label{fig:5}
\end{figure}

The wavelength deviation leak on the pupil plane has a flat wavefront. 
Thus, the leak focuses on the focal plane with the same distribution function profile as on-axis point sources (Figure \ref{fig:5}).
This nature encourages the serial combination of another nuller, which effectively erases the on-axis source, after the 1DDLC.
\subsection{Second-stage Nuller}
\begin{figure}[htb]
    \centering
    \includegraphics[width=0.4\linewidth]{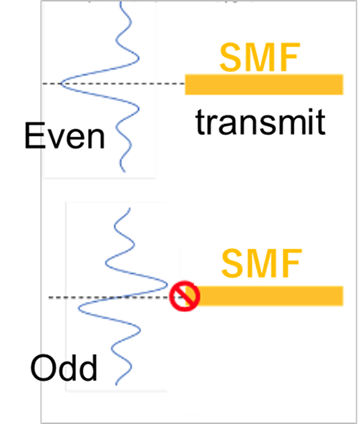}
    \caption{The parity response characteristic of single-mode fibers.}
    \label{fig:6}
\end{figure}

We use the parity fiber nulling \citep[PFN, ][]{2024AJ....167..235I} as the second-stage nuller in the present experiment.
The PFN belongs to fiber-nulling methods, including the vortex fiber nulling \citep{2018ApJ...867..143R}.
The basis of the PFN relies on the parity response characteristic of single-mode fibers (Figure \ref{fig:6}). 
Single-mode fibers transmit only a part of even function components of the amplitude distribution functions of light and no odd function components.
We use this characteristic to distinguish the stellar and planetary light. 

\begin{figure}[htb]
    \centering
    \includegraphics[width=1.0\linewidth]{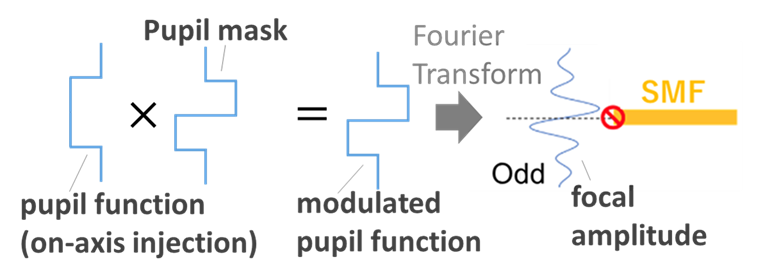}
    \caption{Caption}
    \label{fig:7}
\end{figure}

In the PFN, the light from the on-axis source should emerge as an odd function on the focal plane (the SMF plane).
To modify the parity of the amplitude spread functions of light, we use a mathematical property that the Fourier transform between the pupil and focal planes preserves the parity (Figure \ref{fig:7}).
We can use an odd-function pupil mask (Lyot-plane mask) to modify the light amplitude from the on-axis light source (originally even function) into an odd function.
This modification leads to SMF's null of the stellar leak and transmittance of a part of the planetary light.
\subsection{Combined Nuller}
\begin{figure*}[htb]
    \centering
    \includegraphics[width=0.9\linewidth]{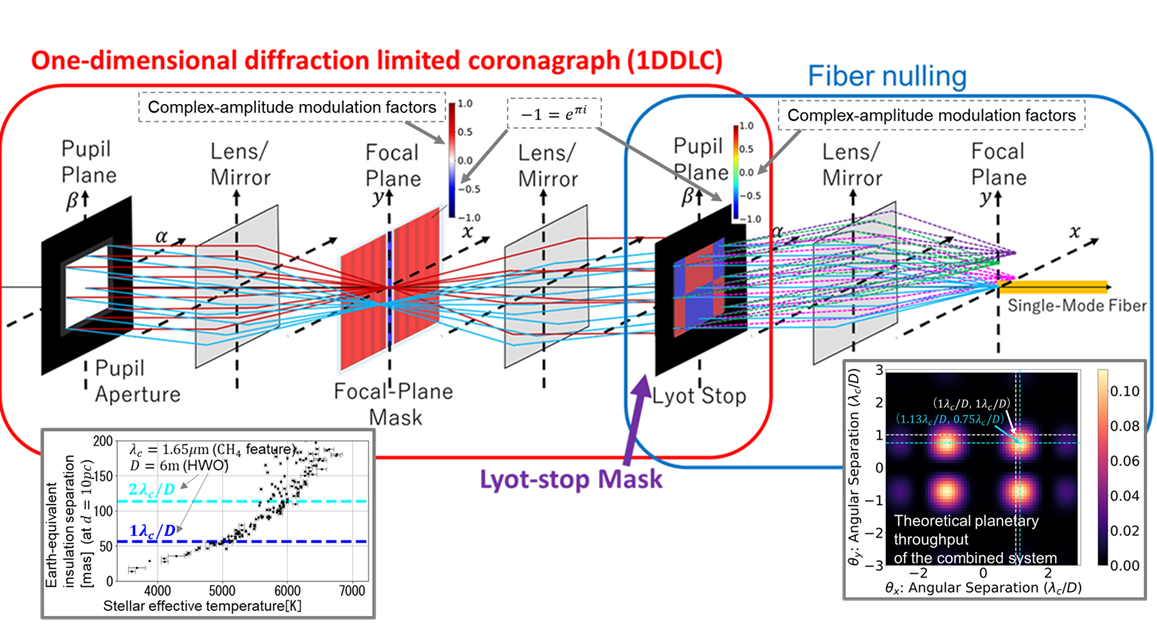}
    \caption{The concept of the combined nuller system.}
    \label{fig:8}
\end{figure*}

\begin{figure*}[htb]
    \centering
    \includegraphics[width=1.0\linewidth]{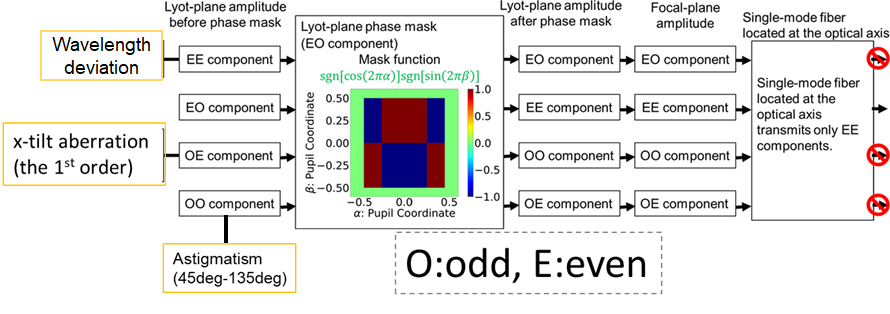}
    \caption{The design strategy for a robust nulling of the stellar light using the combined nuller.}
    \label{fig:9}
\end{figure*}

We combine the first and the second nuller so that the Lyot stop (or its optical conjugate) of the 1DDLC matches the entrance pupil of the PFN (Figure \ref{fig:8}).
The Lyot-stop plane has a phase mask with the following (complex) amplitude modification function $M_{L}(\alpha,\beta)$: 
\begin{equation}
    M_{L}(\alpha,\beta)=\frac{\cos\left(2\pi \alpha\right)}{\left|\cos\left(2\pi \alpha\right)\right|}\frac{\sin\left(2\pi \beta\right)}{\left|\sin\left(2\pi \beta\right)\right|},
\end{equation}
where the symbols $(\alpha,\beta)$ denote the pupil Cartesian coordinates normalized by the pupil rectangular diameters (color maps in Figures \ref{fig:8} and \ref{fig:9}). 

This Lyot-stop mask functions as a diffraction grating with an extremely small number of spatial modulation waves, resulting in diffraction peaks divided into four peaks 1-$\lambda/D$ separated from the optical axis.
Hence, the theoretical planetary throughput of the system shows the dependency on the planetary angular separation, shown in the bottom right panel of Figure \ref{fig:8}.
The present system has a planetary throughput for 1-$\lambda/D$ separated planets, which may be complementary to other coronagraphic systems considered for the HWO.
When considering observations with near-infrared wavelengths of light, the Earth-equivalent-insolation separation angles come close to diffraction-limited separation angles (the bottom left panel of Figure \ref{fig:8}). 
In this near-infrared coronagraph observation, we can utilize the information of the planetary location from the previous visible coronagraph observation, which results in a long exposure time for every single planetary object.

We designed the two-dimensional Lyot-plane mask by considering the robustness to the optical aberrations in addition to the wavelength-deviation leak of the first-stage 1DDLC.
Figure \ref{fig:9} shows the design strategy.
Since we have two dimensions in the focal and pupil planes, we can divide the amplitude spread function into four components: (Even, Even), (Even, Odd), (Odd, Even), and (Odd, Odd) with respect to the first and second coordinates.
Single-mode fibers transmit only (Even, Even) components.
The Lyot-plane phase mask properly modulates the parity of the amplitude spread function so that no leak due to the low-order wavefront aberration goes through the single-mode fiber. 
\section{Experiment}
\begin{figure*}[htb]
    \centering
    \includegraphics[width=0.8\linewidth]{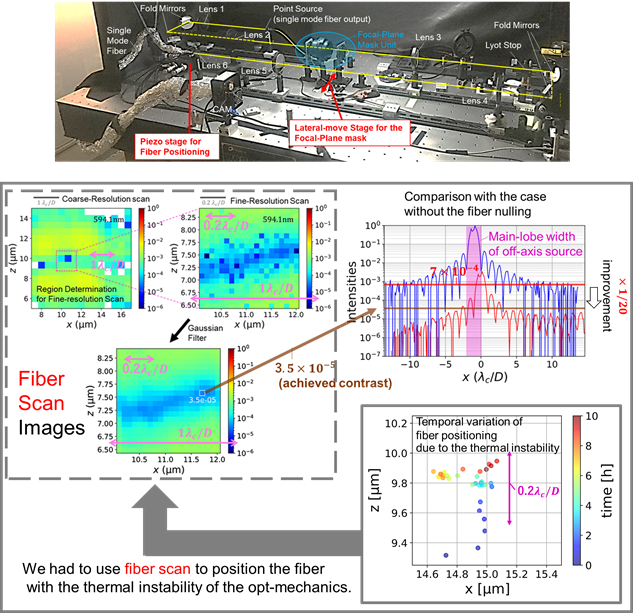}
    \caption{Picture of the experimental setup (top panel) and result (bottom panel) for the wavelength 6.5\% longer than the design-center wavelength (to see the broadband performance). This experiment includes no wavefront correction.}
    \label{fig:10}
\end{figure*}

We have conducted an experiment to validate the concept of the combined nuller. 
The top panel of Figure \ref{fig:10} shows a picture of the experimental setup.
Our previous publications \citet{2023PASP..135f4502I} and \citet{2025PASP..137h4401I} describe a detailed method of the experiment. 
Figure \ref{fig:1} displays the aperture used in the experiment.
Figure \ref{fig:3} shows the focal-plane mask unit used in the 1st-stage 1DDLC.

The size of the point-spread function just before the single-mode fiber must be approximately equal to the mode-field diameter of the single-mode fiber (about a few $\mu m$). Therefore, the experiment requires sub-micron precision in fiber positioning to demonstrate high contrast-mitigation ability.
To achieve precise fiber positioning against thermal instability (the bottom right of the bottom panel of Figure \ref{fig:10}) in the testbed environment, we used a fiber-scan method. This method utilizes a closed-loop controlled piezo stage.
The left of the bottom panel of Figure \ref{fig:10} shows the fiber-scan image, including coarse-resolution scan (preparation for fine-resolution scan) and fine-resolution scan. 

The experiment focused on the contrast-mitigation ability for the wavelength 6.5\% longer than the design-center wavelength (to see the broadband performance). This experiment includes no wavefront correction.
As a result, we confirmed a contrast-mitigation ability of $3.5\times10^{-5}$ for the wavelength 6.5-\% longer than the design-center wavelength.
The value means 20 $\times$ improvement from the preparation experiment for only the 1DDLC without the fiber nulling.
The absence of the wavefront correction in the testbed limits the currently demonstrated contrast mitigation.
\section{Conclusion}
We have reported an experimental verification of a serially combined nuller after theoretical reviews. 
We demonstrated a contrast-mitigation ability of $3.5\times10^{-5}$ for the wavelength 6.5-\% longer than the design-center wavelength (20 $\times$ improvement from the case with only 1DDLC).
The combined nuller theoretically has the ability to null the double stars simultaneously for a wide spectral band. In addition, it has diffraction-limited ($\sim$1$\lambda/D$) inner working angles. Combined nullers, such as the nuller in this paper, will be able to complement the currently considered coronagraph instruments for the HWO.


\bibliography{author}

\end{document}